\documentclass[11pt]{article}
\usepackage{amsmath,amsthm,amssymb,enumerate}
\usepackage{graphicx}
\usepackage{lscape}
\usepackage[english]{babel}
\usepackage[noend]{algpseudocode}
\usepackage[dvipsnames]{xcolor}
\usepackage{float}
\usepackage{tikz}
\usetikzlibrary{calc}
\usepackage[shortlabels]{enumitem}
\usepackage{amsthm}
\usepackage{algorithm}
\usepackage{algpseudocode}
\usepackage{lscape}

\title{\textbf{Intractable Group-theoretic Problems Around Zero-knowledge Proofs}
}
\date{}
\author{Cansu Betin Onur\thanks{EMail: cbetin@metu.edu.tr}\\
 Institute of Applied Mathematics\\Middle East Technical University\\06800 Ankara Turkey}

\begin{document}
\maketitle

\begin{abstract}
While the amount of data produced and accumulated continues to advance at unprecedented rates,  protection and concealment of data increase its prominence as a field of scientific study that requires more action. It is essential to protect privacy-sensitive data at every phase; at rest, at run, and while computations are executed on data. The zero-knowledge proof (ZKP) schemes are a cryptographic tool toward this aim. ZKP allows a party to securely ensure the data's authenticity and precision without revealing confidential or privacy-sensitive information during communication or computation. The power of zero-knowledge protocols is based on intractable problems. There is a requirement to design more secure and efficient zero-knowledge proofs. This demand raises the necessity of determining appropriate intractable problems to develop novel ZKP schemes. In this paper, we present a brief outline of ZKP schemes, the connection of these structures to group-theoretic intractable problems, and annotate a list of intractable problems in group theory that can be employed to devise new ZKP schemes.
 
\end{abstract}

\newpage

\tableofcontents

\section{Introduction}

Privacy and security of data are of vital importance. While storing, processing, or transmitting privacy-sensitive data, it is compulsory to guard it at every stage. The General Data Protection Regulation (GDPR) that came into force in Europe on May 25, 2018, imposes strict requirements for the privacy of data. As a consequence of GDPR, privacy has evolved as a fundamental privilege of a European citizen. When sharing privacy-sensitive personal data, it must be anonymized or encrypted. In various applications such as financial, health, or demographic data analysis where data must be analyzed by and transferred to diverse analytics entities,  strict regulations have to be abided by the principals. Even for conducting research on the data, it has to be shielded to ensure the fundamental right of privacy. Data that crosses borders is much more complex because of the heterogeneity of laws and regulations of countries. Cross-border interoperability requires adaptation of privacy measures implemented based on the specific policies of countries the data crosses. This challenge is called data-sharing problem, and it can be solved even without communicating data in some applications.

Zero-knowledge proof (ZKP)  is a technology that can be extensively employed for preserving the privacy of data and its owner. In a data-based world, privacy can be maintained by ZKP systems by helping principals affirm the veracity of information without disclosing the data. Then, the data-sharing challenge can be managed to an extent with zero-knowledge proofs. Verifying data without revealing them can be employed in many applications, such as identification of a user, secret key agreement, digital signatures, and anonymous credentials. The new threat of the development of quantum computers, the absence of interoperable standards, and the associated computational costs are the barriers to broader adoption of ZKP in real-world applications. 
Classical cryptographic primitives will be insecure after quantum computers which exploit the quantum mechanical phenomenon to solve intractable problems that cannot be solved using conventional computers. If a large-scale quantum computer is built, the intractable problems that are employed for constructing security protocols will be efficiently solved by that quantum computer, and the security protocols will be broken. The power of zero-knowledge protocols is largely based on a few hard-to-solve problems such as discrete logarithm problem or factoring large numbers that we present in Section~\ref{sect:GroupTheoreticProblems}. Since these commonly-used problems can be solved efficiently by quantum computers, the security protocols relying on them have to be revised. We need secure zero-knowledge protocols without leaking any secret information. This demand brings about the requirement of determining suitable intractable problems to cannot be efficiently solved by quantum computers. In 1991, Goldreich, Micali, and Wigderson showed that any NP problem could be used to implement a zero-knowledge proof system \cite{Goldreich1991}. In this paper, we investigate nominee group theoretic algorithmic problems for constructing a secure (and hopefully efficient) zero-knowledge proof protocol.

We organized the paper as follows. We give a background on zero-knowledge proof systems and their relation with group theory in Section 2.  Group-theoretic algorithmic problems are presented and discussed with their complexity classes in Section 3. We overview the basic decision problems defined on groups; word, conjugacy, and isomorphism problems with their generalizations such as subgroup membership, order, power, root, Knapsack types (subset sum, knapsack, submonoid membership), twisted conjugacy, decomposition, double coset membership, factorization, Diffie-Hellman-like conjugacy, power conjugacy, intersection problems (subgroup intersection, coset intersection, set-wise stabilizer). The problems, centralizer, centralizer in another subgroup, set transporter, and restricted graph automorphism, which we consider deserving to be mentioned, are also provided in this section. The intriguing problems of Post's correspondence, endomorphism, weight, subgroup distance, hidden subgroup, and hidden shift in cryptographic structure design are also examined in this section. Eventually, Section 4 concludes the paper.

\section{Zero Knowledge Proof Systems}

We present an overview of zero-knowledge systems in this section. The consideration of group-theoretic problems in the construction of zero-knowledge proofs is intuitively presented.

\subsection{Basic Concept}

 \textbf{Zero-knowledge proofs } (ZKP) are cryptographic constructions where the prover may convince the verifier that a certain statement is valid (true) without revealing any extra information about that knowledge~\cite{Bootle2016}. ZKP schemes have to satisfy the following requirements:
\begin{itemize}
\item \textit{Completeness}: When a statement is TRUE, the prover convinces the verifier.
\item \textit{Soundness}: When a statement is FALSE, a deceitful (dishonest) prover cannot convince the verifier.
\item \textit{Zero-knowledge}: The verifier does not learn any piece of information by running the ZKP scheme other than the fact that the statement is TRUE.
\end{itemize}

The zero-knowledge property, the verifier will not gain any extra information from the interaction, is formulated employing a simulator (a probabilistic polynomial-time algorithm). The simulation paradigm \cite{oded2001foundations} hypothesizes that ``whatever a principal (prover or verifier) can accomplish by itself cannot be considered a gain from interaction with the outside." The output of the simulator and the verifier are to be indistinguishable from each other.

For practical reasons, researchers relax the definition of zero-knowledge by permitting the simulator to fail. Primarily there are three variants of zero knowledge definition; {\bf perfect} (no leaked information), {\bf statistical} (some information may leak to the verifier, however, it is a negligible amount independent of the computational resources the verifier) and {\bf computational} (the amount of information leaked is negligible if the verifier is a probabilistic polynomial-time Turing machine). In other words, it is not possible to notice the difference in the outputs of the verifier and the simulator from an information-theoretic viewpoint in perfect and statistical (a.k.a., almost-perfect) zero knowledge. Albeit, this distinction cannot be expressed by any computationally efficient procedure in computational zero-knowledge. A zero-knowledge protocol without any of these attributes is the most general class of computational zero-knowledge.

\subsection{A Brief History of ZKP}

ZKP schemes are a technology that can be vastly employed for privacy preservation and data ownership. Goldwasser, Micali, and Rackoff introduced ZKP in \cite{goldwasser1989knowledge}.  ZKP schemes are originally designed to be interactive; some messages are exchanged between the prover and the verifier to run the ZKP protocol. Efficiency concerns lead to a variant of ZKP that require no interaction, namely non-interactive zero-knowledge proofs (NIZKP). In this case, the prover only conveys a proof to the verifier and parties do not have to exchange messages online.

Blum et al. established that a common reference string (CRS) shared by a trusted party to the prover and the verifier is sufficient to achieve computational-NIZKP~\cite{blum1988non}. An alternative to CRS is the Random Oracle (RO) model. RO presumes a perfect random function is known to both the prover and the verifier. An interactive ZKP scheme that consists of three steps (commit, challenge, and response) is called a sigma protocol. Fiat and Shamir \cite{fiat1986prove} developed an approach for transforming any sigma protocol into a NIZKP by the use of a cryptographic hash function as a simulator. Goldreich and Oren showed that NIZKP is impossible without any further assumption such as the CRS model or the RO model in the standard model \cite{goldreich1994definitions}.

The most popular protocol which is one of the first examples of zero-knowledge proof is Fiat-Shamir (1987) protocol based on the RSA encryption system; i.e., it relies on the difficulty of the prime factorization problem. Since then, a variety of protocols were developed following the Fiat-Shamir protocol such as the Chaum-Evertse-van de Graaf ~\cite{chaum1987improved}, the Feige-Fiat-Shamir~\cite{feige1988zero},  the Beth~\cite{beth1988fiat}, the Guillou-Quisquater~\cite{guillou1988practical}, the Desmedt~\cite{desmedt1988subliminal} and the Schnorr~\cite{schnorr1991efficient} protocol. All these protocols depend on the computational intractability of  the prime factorization problem or the problem of finding square root in composite modulo with unknown factorization, or discrete logarithm problem. 

The state-of-the-art   ZKP systems are mainly zk-SNARKs \cite{ben2014succinct}, BulletProofs \cite{bunz2018bulletproofs}, zk- STARKs \cite{ben2018scalable} and ``MPC-in- the-head”  \cite{ishai2007zero}. These are non-interactive or can be converted to a non-interactive scheme via the Fiat-Shamir transform. The proposed constructions of SNARKs mostly depend on bilinear maps or lattices and require a trusted setup. The enhancement from SNARKS to STARKS dismissed the requirement for a trusted setup. STARKS rely on collision-resistant hash functions. The STARK-friendly hash function \cite{ben2020stark, canteaut2020report} is the focus of extensive research campaign. A major disadvantage of STARKS is the proof size compared to STARKS. In the MPC-in-the-head approach, a ZKP protocol is designed for NP-relations by using secure multiparty computation (MPC) protocols. Bulletproof depends on the hardness of discrete logarithm problem and has no trusted setup.

\subsection{Group Theory in ZKP}

Goldreich, Micali, and Wigderson showed in 1991 that any NP problem can be employed to construct a zero-knowledge proof scheme \cite{Goldreich1991}, and this development drove ZKP as a prominent technology.  A decision problem is in NP class when a solution instance to the problem is provided, one can efficiently (in polynomial time) check whether or not it is correct. Finding a solution is not efficient for a problem in NP class, but verifying a given solution is efficient. 
 
The studies in the literature connected to the application of group-theoretic intractable problems in ZKP construction either provide complexity analysis and classification of problems, or protocol designs. As an example,  Fenner and Zhang \cite{fenner2005quantum} demonstrated that the group intersection and double coset membership problems can be used to construct an honest-verifier ZKP scheme. They established that these problems are in the statistical zero-knowledge proof class. Arvind and Das categorized several intractable problems in group theory with perfect and statistical zero-knowledge proofs~\cite{arvind2008szk}.  They revealed that the permutation group problems Coset Intersection, Double Coset Membership, and Group Conjugacy are in perfect zero-knowledge class. Sdroievski et al. showed that the hidden subgroup problem has a statistical zero-knowledge proof \cite{SDROIEVSKI2019204}.
 
We enumarate group-theoretic intractable problems as candidates for ZKP protocols in the next section.

\section{Group Theoretic Algorithmic Problems}
\label{sect:GroupTheoreticProblems}
The security of open cryptographic schemes relies laboriously on the difficulty of the underlying intractable problem. The discrete logarithm problem and the prime factorization problem of an integer are the commonly employed problems. Since the subject of this study is group theoretic problems, we will not examine the later problem. 

{\bf Discrete Logarithm Problem (DLP) } Given a finite single-generator (cyclic) group $G=\langle g\rangle $ and an element $h$ from this group,  the question is finding an integer $x$ that satisfies the equality $h=g^x.$ Note that the definition is stated in multiplicative notation. One may prefer using additive notation instead, as in the case $G$ is the group of multiples of a point $P$ on an elliptic curve defined over a finite field.

The first public key cryptosystem, the Diffie-Hellman key exchange, is based on the computational difficulty of discrete logarithm problem (DLP) \cite{diffie1976new}.  Following this protocol, RSA \cite{rivest1978method} and ElGamal \cite{elgamal1985public} cryptosystems were developed in chronological order.  While RSA is based on the difficulty of prime factorization of an integer, ElGamal uses the DLP. Although from a group-theoretic perspective, two cyclic groups of the same order are isomorphic, the existence of an algorithm that efficiently computes the discrete logarithm problem in one of these groups does not imply that there is an algorithm that solves the problem efficiently in the other group. For this reason, the trend to develop efficient cryptographic methods where discrete logarithms are difficult on a cyclic group has been adopted by various researchers.  For example, cyclic subgroups of matrix groups (the problem is not more difficult on these groups than on the multiplicative group of finite fields \cite{menezes1992note}), critical groups of graphs (shown to be unsafe \cite{blackburn2009cryptanalysing, shokrieh2010monodromy}), the group of multiples of a point on an elliptic curve defined over a finite field is used. Among these examples, elliptic curve cryptography (ECC) is an actively studied research area. ECC is differentiated from other candidates by its efficiency in applications and high security in small key sizes.

DLP draws attention in the literature with its structure suitable for creating efficient zero knowledge systems. Discrete-log based ZKP schemes include Groth's protocol \cite{groth2009linear}, the work of Bayer and Groth \cite{bayer2012efficient},  a scheme of Bootle et al. \cite{bootle2016efficient} and Bulletproof \cite{bunz2018bulletproofs}. The variant of DLP involved in the Diffie-Hellman key exchange protocol was later named as Diffie-Hellman Problem.

{\bf Diffie-Hellman Problem (DHP) }  Given a finite cyclic group $G=\langle g\rangle$ and two elements $h=g^x,k=g^y$ from this group where $x$ and $y$ are randomly chosen secret integers. Find the element $s=g^{xy}.$ This problem is sometimes called the Computational Diffie-Hellman problem to distinguish it from its decision version.

{\bf Decision Diffie-Hellman Problem }  Given a cyclic group $G=\langle g\rangle$ of order $n$ and three elements $h,k,s$ from this group satisfying  $h=g^x,k=g^y, s=g^z$ where $x,y,z$ are secret integers. Determining whether $z=xy$ mod $n$ is defined as the decision Diffie-Hellman problem.

Koblitz and Menezes examine several variants of and the reductions between the DHP and DLP in \cite{koblitz2010intractable}. To solve DLP implies the ability to solve DHP. Also over many groups, the DHP is nearly as hard as the DLP  \cite{maurer2000diffie}. In the literature, groups on which cryptographic protocols are defined are called \emph{platform groups}. On a well-chosen platform group, DLP and DHP are the legacy intractable problems.
 
In 1994, Shor \cite{shor1994algorithms} developed a quantum algorithm that solves the factorization and discrete logarithm problems in polynomial time.  After this development, the need of creating alternative methods to design encryption systems has emerged. One of these methods is to use non-commutative cryptographic structures. The complexity of algorithmic problems defined on non-commutative groups has nominated these algebraic structures appropriate tools for cryptographic protocols.

\subsection{Basic Decision Problems Defined on Groups}

A problem that can be answered with a binary result (yes or no) is called {\bf Decision Problem}. Three fundamental decision problems defined on groups were proposed by Max Dehn in 1911~\cite{dehn1911unendliche}. These problems are called word problem, conjugate problem and isomorphism problem. 

\begin{itemize}
\item Word decision problem:
Let a group $ G $ with finite presentation and $ g \in G $ be given. Decide if $ g=1 $.

\item Conjugacy decision problem: 
Given a finitely presented group  $G$ and two elements $a,b$ from $G.$
Determine whether $a$ and $b$ are conjugate in $G$? In other words, is there $ x \in G $ that satisfies $ x^{-1}ax=b $?

\item (Group) Isomorphism decision problem: Let two groups with finite presentations $ G_1, G_2 $ be given. Decide if these groups are isomorphic.
 \end{itemize}

These problems, which are defined on groups with finite presentation, can be generalized to countable groups by enumerating their generators.
In the most broad sense, Dehn’s decision problems given above are undecidable. In other words, they are not algorithmically solvable (Novikov \cite{novikov1958algorithmic}, Adian \cite{adian1957unsolvability})

In the case where the answer to the decision problem is yes, the question of searching for solution instances satisfying the questioned property is called {\bf Search Problem}. Search problems have a large place in the construction of cryptographic structures. Below, the problems are defined in the categories of decision problems and search problems and their place in the literature is presented.

\subsection{Word Problem}

We defined the word decision problem above and stated that there is no algorithm to solve the word problem in general. On the other hand, group classes where the word problem is NP (can be solved in polynomial time with a non-deterministic Turing machine) are quite wide. For instance, finitely generated matrix groups defined on a field, polycyclic groups, finitely-generated metabelian groups, automatic groups, Thompson group, finitely generated subgroups of diagram groups, free Burnside groups $B(m,n)$ for a large enough number $n$ are in this class \cite{birget2002isoperimetric}. There are also groups with finite presentations on which the word problem is NP-complete \cite{sapir2002isoperimetric} or coNP-complete. Birget et al. \cite{birget2002isoperimetric} obtained a group theoretic characterization of NP-problems. They concluded that the word problem on a finitely-generated group $G$ is in NP if and only if the group $G$ can be embedded into a finitely-presented group with polynomial Dehn function.

{\bf Word search problem:}
It is known that a given element $ g \in G $ satisfies $ g=1 .$  Express this element $ g $ as the product of the conjugates of the descriptive relations given in the group presentation.

To the best of our knowledge, there is no ZKP relying on the word problem. The first attempt to establish an encryption system using the word problem was given in 1984 by Magyarik-Wagner \cite{wagner1984public}. This approach is also the first instance of using non-commutative groups. The method used here is based on a problem that is actually easier to secure since it restricts the word problem. In \cite{vasco2004reaction}, Vasco and Steinwandt show that for the proposed group,  the secret key can be discovered by observing the reaction of recipients. On the other hand, it is emphasized that the word problem in groups with finite presentation is an appealing nominee for deriving one-way functions and therefore establishing a public key system. Then, Birget et al. \cite{birget2005public} analyzed the Magyarik-Wagner system and proposed a new public-key encryption system inspired by it. In the new method, the deciphering phase uses the group actions on the words.

One of the early attempts to use the problem was to obtain a one-way function by repeatedly adding new relations to the word produced by the two elements~\cite{do1988public}. This method has remained cumbersome in practice.

The system proposed by Garzon and Zalcstein used the word problem on Grigorchuk groups. It is similar to the Magyarik-Wagner system. The security of this example, which is the first application of the Grigorchuk groups in cryptography, is known to be weak~\cite{petrides2003cryptanalysis}.

Panagopoulos \cite{panagopoulos2010secret} has developed a $(t,n)$-threshold secret sharing scheme using group presentations and the word problem. A $(t,n)$-threshold secret sharing is a method of distributing a confidential information to $n$ participants in a way that while any $ t $ participants can reach the secret, less than $ t $ participants cannot. In the first part of the two-stage scheme, the confidential information  to be used in solving the word problem, such as the defining relations of the group, is distributed to the participants over the secure channel. In the second stage, the actual secret is conveyed over open channels. The prominent difference of this method from the previously developed $ (t,n) $-threshold secret sharing schemes is that the real secret does not have to be transmitted over the secure channel. Habeeb et al. \cite{habeeb2012secret} make this method more practical and gave schemes for sharing confidential information based on group presentations and word problem.

\subsection{Generalizations of the Word Problem}

Various generalizations of the word problem have been presented in the literature. The main ones are subgroup membership problem, order problem, power problem and root problem. 

 \begin{itemize}
    
\item {\bf The Subgroup Membership Problem: }  \\
Given a group $ G ,$ an element $ g \in G $ and a finitely-generated subgroup $ H=\langle h_1, \dots , h_k \rangle $ of $G$.\\
Decision problem: Decide whether $g$ is an element of $H$.\\
Search problem: It is known that $ g \in H $. Express $g$ as a product of powers of generators of $H.$

This problem is also known as {\bf the uniform generalized word problem} or {\bf occurrence problem}.

\item {\bf The Order Problem: }\\
Given a group $G$ in terms of generators and defining relations and an element $ g \in G.$\\
Decision problem: Determine whether $g$ has a finite order.\\
Search problem: It is known that $ g $ has a finite order. Find the order of $g$.

\item {\bf The Power Problem: } \\ 
Given a group $G$ and two elements $x,y \in G.$\\
Decision problem: Determine whether there exists an integer $n$ that satisfies $y=x^n?$ \\
Search problem: It is known that there is an integer $n$ that satisfies $y=x^n$. Find a value of $n$ that satisfies this equation.

The power problem is a special case of the subgroup membership problem. This phenomenon is straightforward when the subgroup is taken as $ H=\langle x \rangle $.

\item {\bf The Root Problem: } \\ 
If the pair $ a, b $ given from a group $ G $ that satisfies $ b = a^n $ for a natural number $ n $, then the element $ a $ is said to be the $n^{\text{th}}$ root of $ b $. Given a group $G,$ an element $ b \in G $ and a natural number $ n .$\\
Decision problem:  Is there an element $ a $ in the group $ G $ that satisfies $ b=a^n $?\\
Search problem: It is known that such element exists. Find an element $ a $ in group $ G $ that satisfies $ b=a^n $.
\end{itemize}

The word problem can be reduced from the order problem by considering the order of the element as one. Again, taking the element $x=1$ in the power problem and choosing $n=0$ in the root problem, one gets the word problem \cite{lipschutz1971groups}. These generalizations are computationally harder than the word problem \cite{mikhailova1966occurrence}, \cite{mccool1970unsolvable}, \cite{lipschutz1971groups}.

The well-known Schreier–Sims\cite{sims1970computational} algorithm solves the subgroup membership problem for permutation groups in polynomial time. Whereas it is undecidable for $F_2\times F_2$ direct product of rank 2 free groups  \cite{mikhailova1966occurrence}.

Various authentication schemes, digital signatures proposed relying on the hardness of the root and conjugacy problem on braid groups 
\cite{wang2009signature,anshel2021walnutdsa,sibert2006entity}.  For braids, generically, Cumplido et al. \cite{cumplido2019root} provide an efficient algorithm which computes the $n^{\text{th}}$ root or assures that such root does not exists.

In \cite{sze2011finding} a key exchange protocol is given based on the difficulty of finding the square root in $ 2 $-Engel groups. $ 2 $-Engel groups are nilpotent groups. The difficulty of the $n$th root problem in nilpotent groups is given in \cite{sze2011finding} by Sze et al. As emphasized here, the difficulty of the problem in nilpotent groups comes from the difficulty of embedding the studied group in the group of upper-triangular matrices defined on rational numbers.

A zero knowledge scheme relying on the root problem \cite{sibert2006entity} is mentioned in Section \ref{conj}.

\subsection{Knapsack Type Problems}

The knapsack problem is one of the most well-known NP-complete problems.  The classic knapsack and subset sum problems are discrete optimization problems concerning integers. Myasnikov et al. \cite{myasnikov2015knapsack} generalized these problems to arbitrary groups by calling them knapsack-type problems. They examined the computational complexity of these problems and their variations.

Note that the below-stated problems are indeed variations of the subgroup membership problem.  In each of the below problems, a group $G$ by a generation set $X,$ an element $ a $ and  a subset $S= \{ g_1, g_2, \dots , g_n \} $ from $ G $ are given. The given generating set $X$ is finite or countably infinite. 

\begin{itemize}
\item {\bf The Subset Sum Problem: }\\
Decision problem: Determine whether $ a=g_1^{k_1} \dots g_n^{k_n} $ for some $ k_i \in \{0,1 \}  $ \\
Search problem: It is known that answer of the decision problem is Yes. Find the values of the exponents.

\item {\bf The Knapsack Problem: }   \\ 
This is a one step generalization of the subset sum problem. The exponents are allowed to be chosen from non-negative integers.\\
Decision problem: Determine whether $ a=g_1^{k_1} \dots g_n^{k_n} $ for some $ k_i \in \{0 \}\cup \mathbb{Z}^+  $ \\
Search problem: It is known that such exponents exists. Find the values of the exponents.

\item {\bf The Submonoid Membership Problem: }   \\ 
Decision problem: Determine whether $ a=g_{i_1} \dots g_{i_s} $ for some $g_{i_k} \in S , s \in \mathbb{N}$ with repetition allowed. \\
Search problem: It is known that there exists such elements. Find values of $g_{i_k}$'s that satisfies this equation. Note that if the given group $G$ is Abelian, the last two problems are equivalent.

Uncited complexity results presented in this section are obtained from \cite{myasnikov2015knapsack}.  The classical Subset Sum Problem (SSP) appear in the group-theoretic setting when $G=\mathbb{Z}$, denoted by $SSP(\mathbb{Z})$,  and in this case depending on the chosen generating set, the problem is pseudo-polynomial. $SSP(\mathbb{Z})$ is in the class {\bf P} for generating set $X=\{ 1   \}$ and, on the other hand, it is NP-complete when $X=\{ 2^n | \ n\in \mathbb{N}   \}$ is choosen. 

In the non-commutative case, $SSP(G)$ is more complex apart from some particular situations. It is NP-complete for a wide range of groups. For example, the Baumslag–Solitar metabelian groups $B(1,p)$ for $p\geq 2$ and $GB=\langle a,s,t | [a,a^t]=1, [s,t]=1, a^s=aa^t  \rangle$ are finitely presented groups having very simple algebraic structures and $SSP(G)$ is NP-complete for these groups. Furthermore, for finitely generated groups (but not finitely presented) seen as competent in terms of calculation such as free metabelian groups of finite rank $r \geq 2,$ for wreath products $A wr B$ where $A,B$ are finitely generated infinite abelian groups, in particular, for $\mathbb{Z} wr \mathbb{Z}$ the problem $SSP(G)$ is NP-complete.  Finitely presented groups are required. It is noted in \cite{myasnikov2015knapsack} that ``SSP(G) is NP-hard if it is NP-hard in some finitely generated subgroup of G." This gives a method to obtain assorted finitely presented groups on which $SSP(G)$ is NP-complete.  Intuitively, the idea is to embed these groups into finitely-presented metabelian groups.   

In \cite{myasnikov2015knapsack}, optimization and bounded versions of above problems are also defined. We are investigating problems with high computational difficulty. To the best of our knowledge, the difficulty of these problems has not yet been proven to be hard in any particular group.  In \cite{myasnikov2015knapsack}, it is shown that the subset sum optimization problems (SSOP1, SSOP2), the knapsack optimization problems (KOP, KOP1, KOP2), bounded knapsack problem together with  $SSP(G),$ search $SSP(G),$ $KP(G),$ search $KP(G),$  and below given $BSMP(G)$ in class $P$  for hyperbolic groups. Also, $SSP(G)$ is known to be in the class  {\bf P} if the base group $G$ is a finitely generated nilpotent group. On the other hand, $G$ is non-virtually nilpotent polycyclic then $SSP(G)$ is NP-complete\cite{nikolaev2018subset}.

\item {\bf The Bounded Submonoid Membership Problem (BSMP): }   \\ 
Decision problem: An element $ a $ and a subset $S= \{ g_1, g_2, \dots , g_n \} $ from a group $ G $ and
 $1^m \in \mathbb{N}$ (in unary) are given. Decide whether $ a=g_{i_1} \dots g_{i_s} $ for some $g_{i_k} \in S  $ and $s \leq m.$ \\
BSMP(G) is NP-hard if $G$ contains an isomorphic copy of $F_2 \times F_2.$ Weaken the group condition, as a monoid example, for integer $n\times n$ matrices with respect to multiplication BSMP is average-case NP-complete whenever $n \geq 20$ \cite{venkatesan1992average}.

\end{itemize}

\subsection{The Conjugacy Problem}\label{conj}

In the conjugacy decision problem, one is asked to determine whether the given elements $a,b$ from a group $G$ are conjugate.
Notice that if there exists an algorithm that can decide the conjugacy problem, then that algorithm decides whether a given element $g$ is conjugate to identity. In other words, solving the conjugacy problem implies solving the word problem. Therefore, we deal with this problem under the restriction that the word problem is solvable in the considered group. 
A  computably presented group with an NP-complete conjugacy problem is given in \cite{miasnikov2017computational}.

Let us state the search version of the problem.

{\bf The Conjugacy Search Problem:  } 
Let it be known that in a given group $ G $, the elements $ a, b $ are conjugates of each other. Find an element $ x \in G $ that satisfies the equality $b= xax^{-1} $.

The computational difficulty of this problem in appropriately selected groups has enabled it to be used in various public-key protocols.  The most popular platform used in the conjugacy problem is the braid group. Braid groups are non-commutative infinite groups that arise naturally from geometric braids. While the group operations on it can be easily calculated, the difficulty of various algorithmic problems such as the conjugacy search problem makes braid groups attractive.

The idea to implement the braid group as a platform for cryptosystems was introduced in 1999 by Ansel-Ansel-Goldfeld. Then, various cryptographic schemes such as Ko-Lee and Sibert-Dehornoy-Girault used the conjugacy search problem on braid groups.

The Ansel-Ansel-Goldfeld (AAG) key exchange protocol is based on the conjugacy search problem on non-commutative groups \cite{anshel1999algebraic}. Although this protocol can be applied to any non-commutative group, the authors specifically emphasize that braid groups are promising. In the proposed scheme, to exchange information between Alice and Bob over the public channel, the word problem must be solved efficiently. The security of this scheme is based on the difficulty of solving the system of conjugates $x u_1 x^{-1} = v_1,  \dots,  x u_m x^{-1} = v_m $  given from the group simultaneously.

In 2019, Grigorchuk and Grigoriev proposed some automaton groups, where the word problem is in polynomial time complexity and it is difficult to solve their conjugate systems, as the platform group for the AAG scheme \cite{grigorchuk2019key}. These are the Grigorchuk group, Basilica group, Universal Grigorchuk group, Three Towers Hanoi group, affine group $Aff_4(\mathbb{Z})$ and certain subgroups of $Aff_6(\mathbb{Z})$.

Ko, Lee et al. have proposed a key agreement protocol based on the conjugacy search problem on \cite{ko2000new} braid groups. In the following process, Shpilrain and Ushakov \cite{shpilrain2006conjugacy} showed that in this protocol an adversary can obtain the common secret key by solving the decomposition problem, which is a comparatively easier problem. A heuristic attack solving the  decomposition problem in the Ko-Lee protocol has been developed in \cite{myasnikov2005practical}, which is much more efficient than any known attack based on solving the conjugacy search problem.

Sibert, Dehornoy, and Girault have proposed three authentication schemes designed for braid groups \cite{sibert2006entity}. The first one is a two-step (challenge-response) authentication scheme based on the (Diffie-Hellman-Like) conjugacy problem, while the other two are authentication schemes based on the conjugacy search problem, which repeats a three-step process several times to guarantee the required level of security. The root finding problem is also used in Scheme III. The authors demonstrated that these schemes they designed are zero-knowledge proofs. On the other hand, Mosina and Ushakov \cite{mosina2010mean} show that Scheme II given here is practically not a computational zero-knowledge proof.

An algorithmic solution to the conjugacy decision and search problems in braid groups was first given by Garside \cite{garside1969braid}. Although this algorithm is not efficient, it has paved the way to the solution to this problem. Then, many deterministic algorithms have been developed to solve the conjugacy search problem in braid groups \cite{birman1998new, elrifai1994algorithms, franco2003conjugacy, gebhardt2005new}. In addition, in \cite{cheon2003polynomial}, a polynomial time algorithm is given for the Diffie-Hellman type conjugacy search problem in braid groups, and an efficient solution to this problem under certain conditions is proposed in \cite{lee2003cryptanalysis}. Apart from that, notable success has been achieved in various length-based or heuristic attacks \cite{garber2002length, hofheinz2003practical}. However, braid groups are still one of the alternatives with potential advantages against quantum attacks.

\subsection{The Generalizations of The Conjugacy Problem}
As in the word problem, many variations of the conjugacy problem have been defined and studied. The main variants of these problems, which are frequently used in cryptology, are presented below.

 \begin{itemize}
\item {\bf The Twisted Conjugacy Problem(TCP): }   \\ 
Given a group $G,$ an endomorphism  $\varphi $ defined on $G$ and given two elements $ a, b \in G$.

Decision problem: Determine whether the equality $ b= x^{-1} a \varphi(x) $ holds for some $ x \in G $.\\
Search problem: It is known that the answer to the decision problem is Yes. Find the element $x$.

This problem is a generalization of the conjugacy problem. When the function $\varphi=id$ is taken, the conjugacy problem is obtained.
 The problem is decidable for a polycyclic groups\cite{roman2010twisted}, for automorphisms of finitely generated free groups \cite{bogopolski2006conjugacy}, for closed surface groups \cite{preaux2016conjugacy} etc. The set of twisted conjugates of an element in a group is called the twisted conjugacy (Reidemeister) class.
The twisted conjugacy classes are based on the Nielsen-Reidemeister fixed point theory. We can also state the problem as: \\
Decision TCP: Is the element $ b $ in the Reidemeister class of $ a $? \\
Search TCP: Find a value $ x $ that allows $b$ to belong to this class\\

In \cite{shpilrain2008authentication} an iterative three-pass authentication scheme is proposed, such as the Fiat-Shamir protocol, based on the difficulty of the twisted conjugacy problem. Here, they named the problem as {\bf the double-twisted conjugacy problem} and defined it for endomorphism pairs as follows:\\

Given a group $G,$ two endomorphisms  $\varphi, \psi $ on $G$ and two elements $ a,b\in G$.\\
Decision problem: Determine if the equality $ b= \psi (x^{-1} ) a \varphi(x) $ holds for some $ x \in G $.\\
Search problem: It is known that such an element exists. Find the element $x$.

The protocol proposed by Ushakov and Shpilrain is given on the semigroup of all $2 \times 2$ matrices over truncated one-variable polynomials over $F_2.$ However, a successful attack was developed for the proposed protocol \cite{grassl2009cryptanalysis}, and then another study \cite{gornova2015cryptanalysis} showed that this protocol is vulnerable in both theoretical and practical terms.

\item {\bf The Decomposition Problem: }   \\ 
Let two elements $ a, b $ and two subsets $ H, K $ from a group $ G $ be given. \\
Decision problem: Is the equality $ hak=b $ holds for some $ h \in H $ and $ k \in K $.\\
Search problem: It is known that $ hak=b $ holds for some $ h \in H $ and $ k \in K.$ Find such elements.

In this problem, when subsets are selected as subgroups, the problem is called the double coset problem.

\item {\bf Double Coset Membership Problem: } \\ 
Let two subgroups $ H, K $ and two elements $ a, b $ from a group $ G $ be given. 
Decision problem: Decide whether the double coset $ HaK $ contains the element $ b.$\\ 
Search problem: It is known that $b\in HaK .$ Find elements $ h \in H $ and $ k \in K $ such that $ hak=b .$

In the double coset problem choose $a=1,$ you get the factorization problem.

\item {\bf The Factorization Problem: }  \\ 
Let two subgroups $ H, K $ and an element $ g$ from a group $ G $ be given. \\
Decision problem: Decide whether the equality $ g=hk $ holds for some $ h \in H $ and $ k \in K $.\\
Search problem: Given that $g$ is an element in the set $HK.$ Find elements $ h \in H $ and $ k \in K $ such that $ g=hk.$

\item {\bf The Number of Factorizations Problem: } \\ 
Let two subgroups $ H, K $ and an element $ g$ from a group $ G $ be given.  Determine the number $n \geq 0$ of distinct factorizations  $ g=hk $ of $g$ where $ h \in H $ and $ k \in K $.\\

\item  {\bf The Diffie-Hellman-Like Conjugacy Problem: }
Given a group $G$ and elements $g,h,k \in G$ such that $h=a^{-1}ga$ and $k=b^{-1}gb$ for some commuting secret elements $a,b \in G.$ Find the element $b^{-1}a^{-1}gab.$

\item  {\bf The Power Conjugacy Problem: }
Two elements $a,b$ in a group $G$ are said to be power-conjugate if there exists integers $n,m$ such that $a^n$ and $b^m$ are conjugate in $G.$
Given a group $G$ and two elements $a,b \in G.$\\
Decision problem: Decide whether $a^n$ and $b^m$ are conjugate in $G$ for some integers $n,m.$\\
Search problem: Given that $a$ and $b$ are power-conjugate in $G.$ Find two integers $n,m$ and an element $x \in G$ such that $b^m=x^{-1}a^nx.$

\subsection{Intersection Problems}

\item Subgroup Intersection Problem:
Given a group $G$ and its two finitely generated subgroups $H, K$ by generating sets. \\
Decision problem: Is the intersection of subgroups $ H $ and $ K $ finitely generated? \\
Search problem: Given that $H \cap K$ is finitely generated. Find a generating set of the intersection.

\item Coset Intersection Problem:
A group $G,$ an element $a \in G$ and generating sets of two subgroups $H,K$ of $G$ are given.\\
Decision problem: Decide whether the intersection $ Ha \cap K $ is empty. \\
Search problem: Given that $Ha \cap K$ is non-empty. Find a generating set of the intersection.

It is easy to see that the factorization problem and the coset intersection problems are equivalent. For the convenience of the reader,  the observation of the equivalence is given: an element $a \in G$  is of the form $a=hk$ for some $ h \in H $ and $ k \in K $ only if  $k=h^{-1}a \in Ha \cap K.$

\subsection{Problems based on group actions and invariants}

\item Setwise Stabilizer Problem:
Let a group $ G $ act on a finite set $ \Omega $. For a given subset $\Delta$ of $\Omega,$
 find the set $Stab_G(\Delta)=\lbrace g\in G |  \Delta^g=\Delta \rbrace $.

\item Set Transporter Problem: 
Let a group $ G $ act on the set $ \Omega $. 
Determine whether there is an element $ g \in G $ that satisfies $ \Delta_1^g=\Delta_2$ for the given subsets $ \Delta_1, \Delta_2 $ of $ \Omega. $

\item Centralizer Problem: 
 Given a group $ G $ with an element $x$ and a subgroup $ H$. Find the centralizer $ C_H(x)=\lbrace h\in H \ | \ hx=xh \rbrace$ of the element $ x $ in $ H $.

\item Centralizer in Another Subgroup Problem: 
 Let a group $ G,$ and two subgroups $ H, K \leq G $ be given. Find a generating set for the centralizer
  $ C_H(K)=\lbrace h\in H \ | \ hk=kh \text{ for all } k\in K \rbrace  .$

\item Restricted Graph Automorphism:
Given a graph $G=(V, E)$ and a permutation group $H \leq Sym(V).$ Find a generating set for $H \cap Aut(G).$

In \cite{hoffmann1982subcomplete}, Hoffmann presented eight group theoretic problems defined on permutation groups. These are double coset membership, group factorization, number of factorizations, coset intersection, subgroup intersection, setwise stabilizer,  centralizer in another subgroup, and restricted graph automorphism problem. We give these problems above in a generic sense. The problems that Hoffmann introduced are translated from combinatorial problems including graph isomorphism. It is given that all of these problems are {\bf NP}. He also proved that finding generating set for the intersection of the commuting permutation group given by generating set is the class {\bf NP $\cap$ coNP }

E.M. Luks stated that with the Karp reduction method, the group isomorphism decision problem on permutation groups in polynomial time can be reduced to the graph isomorphism problem and this problem can be reduced to the set stabilizer problem; 
He also demonstrated the equivalence of various problems in polynomial time over permutation groups named Luk's equivalent class \cite{luks1993permutation}. This class includes the subgroup intersection problem, the set stabilizer problem, the centralizer problem, the set transporter problem, the double coset decision problem, and many more. In 2016, it was shown that these problems can be solved in quasipolynomial $ exp(log n)^{O(1)} $ time \cite{babai2016graph}.

\subsection{The Post's Correspondence Problem in Groups}

Another classic undecidable problem that  Myasnikov et al. \cite{myasnikov2015knapsack} generalized to arbitrary groups is the Post's correspondence problem (PCP)~\cite{post1946variant}. It is known that bounded PCP is {\bf NP}-complete (SR11 in \cite{Garey1978ComputersAI}).
They study this problem together with its variations and examined the computational complexity of these problems.

\begin{itemize}
    \item {\bf The Post's Correspondence Problem in a Group PCP(G): } Given a finitely generated group $G$ and and a finite set of pairs $(g_1,h_1), \dots, (g_n,h_n)$ of elements in $G.$ Determine if there is a word $w(x_1, \dots, x_n)$ in the free group $F(X)$ with basis $X=\{x_1, \dots, x_n  \},$ which is not identity of $G$, such that 
 $w(g_1, \dots, g_n)= w(h_1, \dots, h_n)$ in $G.$
    
    \item  {\bf The non-homogeneous Post's Correspondence Problem in a Group NPCP(G): } Given a finitely generated group $G$ and and a finite set of pairs $(g_1,h_1), \dots, (g_n,h_n), (a_1,a_2), (b_1,b_2)$  of elements in $G.$ Determine if there is a word $w(x_1, \dots, x_n)$ in the free group $F(X)$ with basis $X=\{x_1, \dots, x_n  \},$ such that 
 $a_1 w(g_1, \dots, g_n) b_1= a_2 w(h_1, \dots, h_n) b_2$ in $G.$ Here the pairs $(a_1,a_2), (b_1,b_2)$ are called the constants of the instance.     
\end{itemize}
 
Notice that in NPCP(G), $w$ does not have to be non-identity. As in the original PCP, the problems PCP(G) and NPCP(G) are also considered in the case where the number of pairs given from the group is bounded above. 

Specifically in \cite{myasnikov2015knapsack}, it is given that the bounded NPCP(G) is {\bf NP}-complete in the case that $G$
 is a non-Abelian free group of finite rank. As a corollary, it is given that NPCP(G) is {\bf NP}-hard if $G$ contains a free abelian subgroup $F_2.$

\subsection{The Endomorphism Problem}

Given a group  $G$ and two elements $g,h \in G$.\\
Decision problem: Determine whether there is an endomorphism $\phi$ of $G$ such that $\phi(g)=h$.\\
Search problem: It is known that there is an endomorphism $\phi$ sending $g$ to $h.$ Find $\phi$.

The endomorphism decision problem is unsolvable for a number of groups \cite{roman1977unsolvability,roman1979equations}.  It is observed that \cite{grigoriev2008zero} for well-chosen platform groups, like free metabelian group of rank two, the problem is NP-hard. In \cite{grigoriev2008zero} a general framework for a zero-knowledge authentication scheme in the form Feige-Fiat-Shamir-like construction is given. Here, the idea is to use a hard-to-invert action of a semigroup on a set. In particular, when a free metabelian group is used as the platform  group, this study gives a zero-knowledge proof relying on the hardness of endomorphism problem. Later, Roman'kov and Erofeev \cite{erofeev2012constructing} present a method for constructing a possibly one-way function on a group with the decidable word problem and undecidable endomorphism problem.  As an application, a zero-knowledge authentication protocol is given. 

\subsection{The Group Isomorphism Problem}
The last Dehn's decision problem we discuss is the group isomorphism problem. This problem has been proved to be undecidable in general. On the other hand, when the groups are given in other ways the problem arises with different complexities. For example, if the groups are given in multiplication tables, this problem can be reduced by polynomial reduction to well known {\bf NP}-problem the graph isomorphism problem \cite{miller1979graph}.

\subsection{The Hidden Subgroup Problem}
Let $G$ be a group, $H$ be a subgroup of $G,$ and $S$ be a finite set. A function $ f: G\rightarrow S $ is said to {\bf hide} $ H $  if for any $g_1, g_2 \in G$, the equality $ f(g_1)=f(g_2) $  is true if and only if $ g_1H=g_2H $. 

\textbf{ The Hidden Subgroup Problem - HSP:} 
Given a group $G$ and oracle access to a function  $ f: G\rightarrow S $ which hides a subgroup $H$ of $G.$ Find a generating set for $H.$ 

HSP has interesting instances such as the well-known computational hard problems on which today's cryptographic structures are based: the integer factorization problem, the discrete logarithm problem, the period-finding problem, the graph isomorphism problem, or the shortest vector problem. It appears in most of the quantum algorithms developed to date. For example, Simon's problem is the problem of distinguishing the trivial subgroup from an order two subgroup in the group $Z_2^n$ and can be solved on quantum computers. Shor's algorithm is inspired by Simon's problem. Shor's integer factorization and discrete logarithm quantum algorithm solve a special case of the HSP. Shor's algorithm for integer factorization is equivalent to the Abelian HSP problem. Shor's applications to the graph isomorphism problem and the shortest vector problem on the lattice are equivalent to the HSP problem over other non-Abelian groups. 

There are polynomial-time efficient quantum solutions for the Abelian HSP \cite{hallgren2003hidden}. The solution to the HSP problem in non-Abelian groups will revolutionize the quantum world. Because this solution will also solve the graph isomorphism problem and the shortest vector problem on lattices. As it is known, lattice-based problems that are used in most of the methods proposed by NIST are quantum-safe. If the shortest vector problem on lattices is solved, the quantum security of most solutions submitted to NIST's competition will also be questioned. For this reason, HSP is a well-studied problem among quantum researchers.

The complexity of problems such as HSP can be measured in two ways. The first metric is query complexity. The query complexity of an algorithm is the number of times the function $f$ is evaluated by an oracle, which is developed as a black box. The other measure is computational (or time) complexity. Computational complexity indicates how many times basic operations such as addition and multiplication are performed in the algorithm.  An algorithm developed for solving HSP is accepted as efficient if both query and computational complexity are polynomial in $log(|G|)$ \cite{koiran2007quantum}. 

Ettinger et al. \cite{ettinger2004quantum} gave a quantum algorithm that identifies a hidden subgroup of an arbitrary finite group in polynomial query complexity. However, it has exponential time complexity.
Yet, it is an open problem whether there is an efficient solution for HSP on the arbitrary non-abelian group.  Instances of the non-Abelian HSP that are not known to have efficient quantum algorithms include the dihedral hidden subgroup problem (DHSP) and the symmetric hidden subgroup problem.

Regev\cite{regev2004quantum} gave a solution to the unique shortest vector problem (SVP), which lies in {\bf NP} and {\bf coNP}, under the assumption that there exists an algorithm that solves DHSP by coset sampling. The best result for DHSP is the Kupenberg's quantum algorithm \cite{kuperberg2005subexponential}, which runs in time and query complexity $2^{\emph{O}(\sqrt{log|G|})}$ where $|G|$ is the order of the group. In other words, it is subexponential. The space requirement of this algorithm is also subexponential. Regev \cite{regev2004subexponential} described a modified algorithm whose space requirement is polynomial and running time is still subexponential but slightly higher than Kuperberg’s algorithm.
It is known that an efficient quantum algorithm for the symmetric hidden subgroup problem implies an efficient quantum algorithm for graph isomorphism \cite{boneh1995quantum},\cite{ettinger1999quantum}.

{\bf Hidden Shift Problem:}\\
Given a finite group $ G,$ a set $ X $ and oracle access to two injective functions $ f, g: G \rightarrow X$   with the property that there exists an element $ s \in G $ such that the equality $ g(x) = f(xs) $ holds for all $ x \in G$. Find the secret shift element $s.$ 

The hidden shift problem is also known as {\bf the hidden subgroup problem}. For the abelian group $G=\mathbb{Z}_n$ this problem is equivalent to DHSP. For arbitrary non-Abelian group $G,$ the hidden shift problem on $G$ with secret $s$ gives HSP on the wreath product $G \wr S_2 $ with the hidden subgroup $H=\langle (s,s^{-1},1) \rangle.$  Indeed symmetric groups, the hidden shift, and the hidden subgroup problems are equivalent \cite{hallgren2010limitations}.

\subsection{Problems Driven by a Metric}

It is possible to define different metrics on groups. Natural algorithmic problems on an algebraic structure equipped with a metric are based on the concepts of length and distance. 

Let $d$ be metric on $S_n.$ Corresponding to given metric, the weight of an element $\alpha$ is defined by $w(\alpha)=d(\alpha, e)$ where $e$ is the identity permutation. The distance between a subgroup $H$ of $S_n$ and an element $\alpha$ is defined by $d(\alpha,H)=min \lbrace d(g, h) \  | \ h\in H  \rbrace.$

 {\bf Weight Problems:}\\
Given a generating set of permutation group $G$ where the generators are in the form of products of cycles and a value $k.$
\begin{itemize}
\item  {\bf Weight Problem:}\\ Decide whether there is an element $\alpha \in G$ such that $w(\alpha)=k.$

\item {\bf Maximum Weight Problem:}\\Decide whether $max \lbrace  w(\alpha)\  | \ \alpha\in G  \rbrace \geq k .$

\item {\bf Minimum Weight Problem (MWP):}\\Decide whether $min \lbrace  w(\alpha)\  | \ \alpha\in G-\{ e \} \rbrace  \leq k .$

\end{itemize}

{\bf Subgroup Distance Problem (SDP):}\\
Given a set of elements $\lbrace g, h_1, \dots h_m \rbrace$ from the Symmetric group $ S_n $ and given an integer $k.$ Decide whether the distance between $g$ and the subgroup $H=\langle h_1, \dots h_m \rangle$ is at most $k.$

Notice that SDP and MWP  can be seen as a non-commutative variation of the closest vector problem and the shortest vector problem of integer lattices.

The computational hardness of SDP is studied along a variety of metrics.  In 2006, Pinch proved that under the consideration of Cayley Distance, SDP is NP-Complete [Pinch, 2006]. Later Buchheim et al. 2009 observed that this result is also true for Hamming, $l_p$, Lee's, Kendall's tau, and Ulam's Distance.  Approximately a year later Cameron and Wu, giving a reduction from the weight problem and the minimum weight problem on permutation groups to the corresponding coding problems, show that these problems are NP-complete for the same metrics list extended by $l_\infty$ and movement metrics.
For the case of a maximum weight problem, authors show that the problem is in class $P$ for $l_\infty$ yet it is still NP-complete for Hamming, movement, Cayley, $l_p$, Lee, Kendall’s tau, and Ulam metrics even for elementary abelian 2-groups with small size orbits. We present the named distances in Table~\ref{tab:distances}.

\begin{landscape}
\noindent\begin{table}[H]
    \caption{The distance $d(\alpha,\beta)$ given two permutations $\alpha, \beta$ in $S_n.$ }
    \label{tab:distances}
    \centering
    \begin{tabular}{p{0.15\textwidth}|p{0.8\textwidth}|p{0.45\textwidth}}
    \textbf{Distance} & \textbf{Description} & \textbf{Explanation} \\ \hline
    Cayley
  &    $min\lbrace  t  |    \alpha \tau_1 \tau_2 \dots \tau_t= \beta    \rbrace$ where each $\tau$ is a transpotition   
  & the minimum number of transpositions  transfering $\alpha$ to $\beta$    
    \\ \hline
      
    Hamming 
    & $|\lbrace i | \ \alpha(i) \not= \beta(i)   \rbrace |$  
    &  the number of different entries of $\alpha$ and $\beta$ 
    \\     \hline
    
    $l_p$ 
    & $\sqrt[p]{\sum_{i=1}^{n}  {( \alpha(i) - \beta(i)   )}^p }$ 
    &   
    \\ \hline
   
    $l_{\infty}$  
    & $max\lbrace  |\alpha(i) - \beta(i)| \ | \  1\leq   i   \leq   n  \rbrace$  
    & the maximum value of entry difference 
    \\ \hline

 Lee's 
 & $\sum_{i=1}^{n}  min ( |\alpha(i) - \beta(i)| , n-|\alpha(i) - \beta(i) |   )$ 
 &   \\ \hline

  Kendall's tau 
   &  $min\lbrace  t  |  \  \alpha \tau_1 \tau_2 \dots \tau_t= \beta    \rbrace$ where  $\tau_1, \dots, \tau_t $ adjacent transpositions      
   & 
 the minimum number of adjacent transpositions  transferring $\alpha$ to $\beta$ 
  \\ \hline
 
 Ulam's  
 &  $n-l$   where $l$ is the  length of  the longest increasing subsequence in the sequence $( \alpha\beta^{-1}(1) , \dots, \alpha\beta^{-1}(n) )$      
  &   \\ \hline

Movement
   &  $M(\beta\alpha^{-1})$ where  $M(\pi)=\max \limits_{A\subseteq \lbrace 1, \dots, n \rbrace} 
   \pi(A)-A  $     
   & 
metric induced by movement 
  \\ \hline  
  
\end{tabular}
\end{table}
\end{landscape}

{\bf Fixed-Point-Free Permutation Problem (FPF):} This is a special case of the maximum Hamming weight problem by setting $k=n.$  An element $\alpha$ is fixed-point-free if $fix_{\omega}\alpha=\{ x \in \omega \ | \ x\alpha= x     \}= \emptyset.$ That is $w(\alpha)=n.$ A permutation group $G \leq S_n$ is called fixed-point-free if it contains a fixed-point-free element. FPF is determine whether $G$ is fixed-point-free. As noted above, this problem is NP-complete even for elementary abelian 2-groups with orbit size bounded by 4 \cite{cameron2010complexity}.

\end{itemize}

\section{Conclusion}

Zero knowledge proof (ZKP) schemes are broadly employed cryptographic protocols. The security of ZKP schemes relies on computational hardness of intractable problems. This paper aims at presenting intractable problem candidates in the broad field of group theory. We introduced ZKP systems together with their relationship to group theory and examined intractable group-theoretic problems. The commonly-employed problem in ZKP schemes is the discrete logarithm problem, and its variant, the Diffie-Hellman problem on cyclic groups. Earlier studies show that group intersection, double coset membership, coset intersection, group conjugacy and hidden subgroup problems are also in zero knowledge class. There are also zero knowledge schemes on non-commutative groups relying on the root problem, the conjugacy problem and the endomorphism problem. We believe that some of the problems we reviewed in this paper can be employed to construct efficient ZKP schemes that are less quantum-susceptible than commonly used zero-knowledge schemes.

 \section*{Acknowledgement}
Version 1 and 2 of this paper is partially funded by the NLnet foundation under the MoU number 2021-12-510. Version 3 is partially founded by TUBITAK 2219 under grand agreement 1059B192001003.

\bibliographystyle{IEEEtran}
\bibliography{references}

\end{document}